\documentstyle[twoside,fleqn,epsfig,espcrc2]{article}

\newcommand{\AmS}{{\protect\the\textfont2
  A\kern-.1667em\lower.5ex\hbox{M}\kern-.125emS}}

\title{Air shower simulation using {\sc geant4} 
and commodity parallel computing}

\author{L. A. Anchordoqui\address{Department of Physics, Northeastern  
University, Boston, MA 02115, USA}\thanks{doqui@hepmail.physics.neu.edu}, 
        G. Cooperman\address{College of Computer Science, Northeastern 
        University, Boston, MA 02115, USA}\thanks{gene@ccs.neu.edu},
        V. Grinberg$^b$\thanks{victor@ccs.neu.edu},
        T. P. McCauley$^a$\thanks{mccauley@hepmail.physics.neu.edu}, 
        T. Paul$^a$\thanks{tom.paul@hepmail.physics.neu.edu},
        S. Reucroft$^a$\thanks{stephen.reucroft@cern.ch},  
        J. D. Swain$^a$\thanks{john.swain@cern.ch}, $\,$ and 
        G. Alverson$^a$\thanks{george.alverson@cern.ch}}
       
\begin{document}
\begin{abstract}
We present an evaluation of a simulated  cosmic ray shower, based on
{\sc geant4} and {\sc top-c}, which tracks all the particles in the shower. {\sc top-c} 
(Task Oriented Parallel C) provides a framework for parallel algorithm 
development which makes tractable the problem of following each particle. 
This method is compared with a simulation program which employs the Hillas 
thinning algorithm. 
\end{abstract}
\maketitle

\section{Introduction}

The steeply falling end of the cosmic ray spectrum now extends up to
$\approx 3 \times 10^{20}$ eV (see Ref. \cite{yd} for a recent
survey), three orders of magnitude higher than the highest energy 
achieved by
hadron colliders. Direct measurements using sophisticated
equipment on satellites or high altitude balloons are limited in
detector area and in exposure time. Ground-based detectors with
large apertures make such a low flux detectable after a
magnification effect in the upper atmosphere. Namely, the incident
cosmic radiation interacts with atomic nuclei of the air molecules
and produces extensive air showers which spread out over large
areas. This indirect method of detection bears a number of serious
difficulties in the interpretation of the recorded data. In
particular, since many variables are involved the processes
describing the shower development are intrinsically complicated, 
numerical simulation of the giant cascades has to be
performed. The most important source of fluctuations in  Monte
Carlo simulations such as {\sc corsika} \cite{corsika} and {\sc aires} 
\cite{sergio}  are the depth and
characteristic of the first few interactions, necessarily related
to the quality of our understandying of hadronic collisions.

The parts of the shower model which are based on electromagnetic or weak 
interactions can be calculated with ``good accuracy''. The hadronic 
interaction, 
however, is still subject to large uncertainties. In recent years, many 
models of hadronic interactions have been built around experimental results, 
predominantly of $p\bar{p}$ colliders. Extrapolations to higher energies, 
to small angle processes, and to nucleus--nucleus collisions have been 
performed with varying levels of sophistication. In particular, the algorithms 
of {\sc sibyll} \cite{sibyll} and {\sc qgsjet} \cite{qgsjet} are  
tailored for efficient operation to the highest cosmic ray 
energies.\footnote{It 
is important to stress that the latest version of {\sc dpmjet} was improved 
for  operation up to primary energies of $10^{21}$ eV (per nucleon in he lab. 
frame) \cite{ranft}. However, 
it was not yet efficiently implemented in {\sc aires}, 
nor in {\sc corsika}.} 
The different approaches used in these codes to model the underlying
physics can lead to different results when applied to the same 
data \cite{prdhi}. Hence, considerable systematic uncertainties remain.
In addition, a significant 
deviation was 
recently reported in the predictions of {\sc aires} and {\sc corsika}, even 
when both programs 
invoked the same hadronic interaction model to account for the first 
generation of particles \cite{tere}. These discrepancies could result from 
different
energy cuts for the hadronic interaction routines, or  there may be some 
discrepancies in the sampling techniques implemented in these programs 
to reduce 
the number of generated particles. We, therefore, considered it 
worthwhile to compare the simulation results of {\sc aires} 
(which utilizes the Hillas thinning algorithm \cite{hillas}) to that obtained 
with {\sc geant4}+{\sc top-c} 
\cite{neu6} ({\sc top-c} makes tractable the problem of following each 
particle during the atmospheric cascade).\footnote{{\sc geant4} is the 
latest stage in the development of the 
GEANT software, superseding the earlier FORTRAN versions with a new 
object-oriented approach in C++ \cite{GEANTREF}. {\sc top-c} 
(Task Oriented Parallel C) provides a framework for parallel algorithm 
development \cite{gene}.}

\section{Air shower simulation}

\begin{figure}
\begin{center}
\epsfig{file=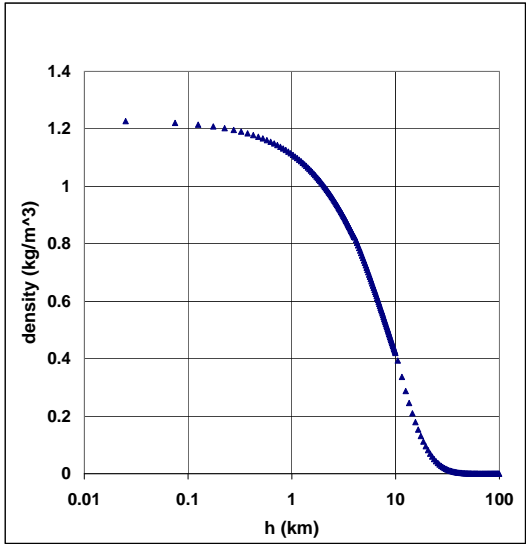,width=7cm,clip=}
\caption{Density of air as a function of the vertical 
altitude above sea level.}
\end{center}
\end{figure}

The most direct way to test the predictions
of {\sc aires} and {\sc geant4}+{\sc top-c} is to study the 
characteristics of the secondaries generated under similar conditions. 
Therefore, 
in order to reduce possible differences caused by the implementation 
of hadronic collisions, we analyze the showers induced by 10 TeV protons. 
Furthermore, we have fixed the 
energy cuts of {\sc geant4} according to the default values of {\sc aires}. 
All shower particles with energies above the following thresholds were 
tracked: 
750 keV for gammas, 900 keV for electrons and positrons, 10 Mev for muons, 
60 MeV for pions and kaons and 120 MeV for nucleons. The particles were 
injected at 
the top of the atmosphere (100 km.a.s.l) and the ground  was located 
at sea level. In the simulations  with  {\sc geant4}+{\sc top-c}, the 
atmosphere was defined by a stack of 230 layers of increasing thickness 
and decreasing 
density (with the height above sea level). 
The geomagnetic field was not taken into account.
The variable density was modeled 
using Linsley's parametrization of the U.S. Standard Atmosphere   
\cite{atmosphere}, leading to the density profile shown in Fig. 1. 
Each point in the figure corresponds to a layer starting with thickness 
50~m at sea 
level and increasing  to  1 km at the top of the atmosphere. 
The above description is consistent with that used in {\sc aires}. 
{\sc qgsjet} was used to model the hadronic processes in the 
simulation with {\sc aires} (we note that at 10 TeV the results obtained 
with {\sc aires}+{\sc sibyll} are almost the same).
 
In Fig. 2 we show preliminary results of the longitudinal development of the 
total number of
secondaries that a 10 TeV proton may produce after cascading in the 
atmosphere. We also show the longitudinal development of different groups 
of secondary particles (we have considered separately $\gamma$, $e^\pm$, 
and $\mu^\pm$). 

At 600 g/cm$^2$  the ratio between the total number of particles produced
by {\sc geant4} and the total number of particles produced by {\sc aires} 
is $\approx 2$. This ratio increases as the shower front gets closer to the 
ground, and roughly reaches the value of 4 at 800 g/cm$^2$.
The main reason for this divergence  comes from a difference in the number 
of charged pions produced during the shower development. In the first 
generation of particles (governed by the hadronic interaction) 
{\sc geant4} produces more charged pions than {\sc aires}. Consequently, 
in the first steps, the muon shower profile developes faster in {\sc geant4} 
than in {\sc aires}. 

The photopion production threshold is at a photon energy  of 145 MeV 
in the proton rest 
frame, with a strongly increasing cross section in the region of the 
$\Delta$ resonances, which decay into neutral and charged pions. With 
increasing energy photon-proton collisions are no longer mediated by 
resonances, but by the diffractive production of the vector mesons, 
$\rho$ and 
$\omega$, or direct multi-pion production. 
We are currently evaluating whether the physical models available in 
{\sc geant4} are sufficient to describe photon-nuclear interactions in the 
energy range of interest. We note that {\sc geant4} provides a framework 
for introducing new physical models if need be. Though our ({\sc geant4}) 
simulation
produces $\gamma$'s above the photopion threshold, we observe no charged 
pions from this process. Therefore, as the shower front approaches the 
ground, the neutral pion decay channel dominates the cascade, yielding a 
huge number of electrons and positrons. As a result, the total number of 
particles is increased, and the position where the shower develops the 
maximum number of particles is shifted to greater depths. 

\section{Final Comment}
The above comparison between {\sc aires} and {\sc geant4} is not yet 
complete. A detailed discussion including the geomagnetic field effects,  
study of the $\gamma p$--interaction, 
and the analysis of different kinds of primaries at ``Auger 
energies'' \cite{auger} will be given in a future publication.

\section*{Acknowledgments}

We have benefited from discussions with Mar\'{\i}a Teresa Dova and Sergio 
Sciutto. This work was supported by CONICET, and the National Science 
Foundation.

\begin{figure}
\begin{center}
\epsfig{file=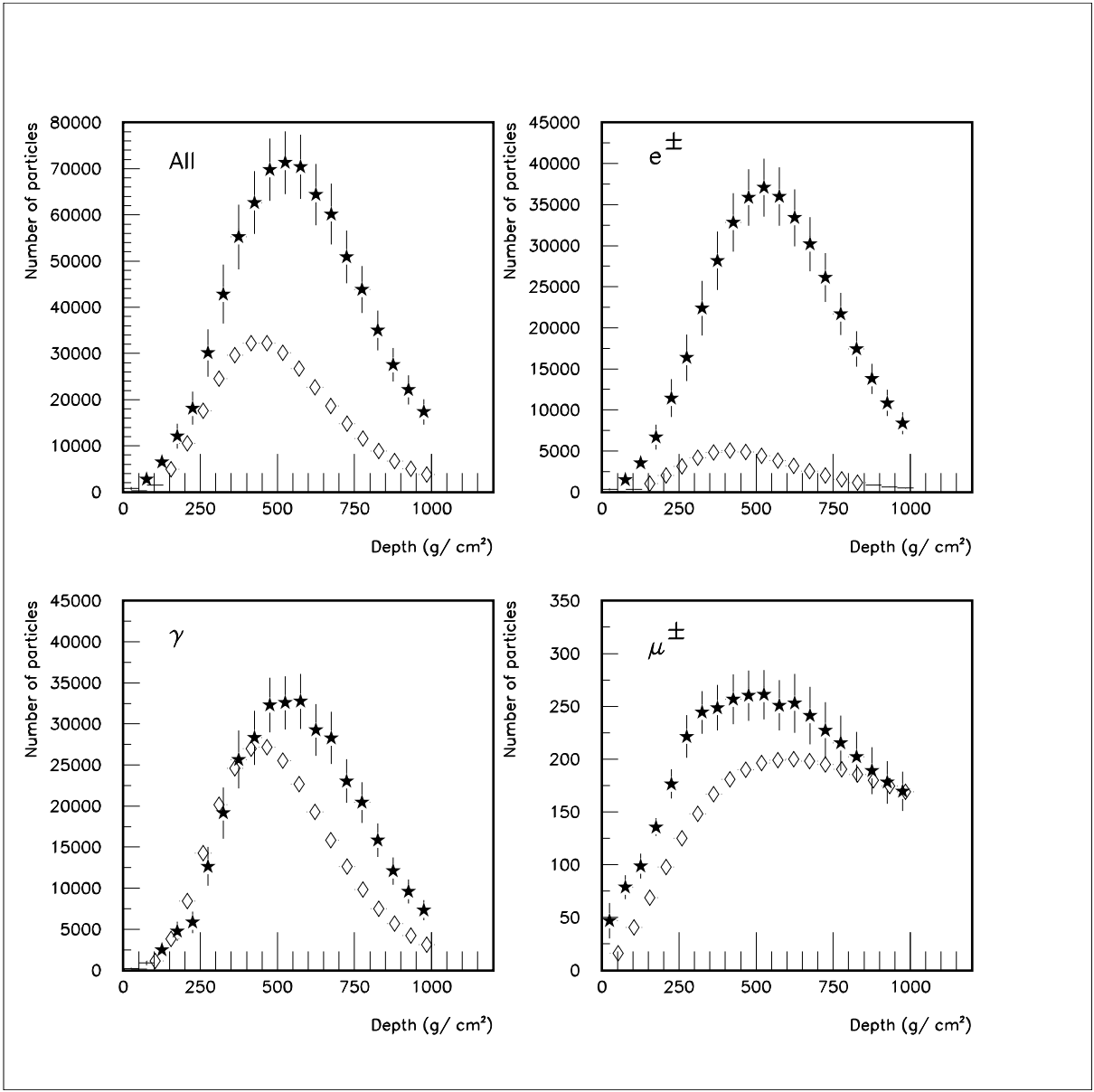,width=16cm,clip=}
\caption{Longitudinal profiles of different secondary species predicted by
{\sc aires} ($\diamond$) and {\sc geant4} ($\star$), for showers induced by 
protons of 10 TeV.}
\end{center}
\end{figure}

\begin{thebibliography}{99}

\bibitem{yd} S. Yoshida and H. Dai, J. Phys. G {\bf 24}, 905 (1998).
\bibitem{corsika} D. Heck {\it et al.}, {\sc corsika}  
{\it (COsmic Ray Simulation for KASCADE)}, FZKA6019 (Forschungszentrum 
Karlsruhe) 1998; updated by D. Heck
and J. Knapp, FZKA6097 (Forschungszentrum Karlsruhe) 1998.

\bibitem{sergio} S. Sciutto, 
{\it Air Shower Simulations with the} {\sc aires} {\it system},
in {\it Proc. XXVI International Cosmic Ray Conference}, (Eds. D.
Kieda, M. Salamon, and B. Dingus, Salt Lake City, Utah, 1999)
vol.1, p.411, [astro-ph/9905185] at http://xxx.lanl.gov.

\bibitem{sibyll} R. S. Fletcher, T. K. Gaisser, P. Lipari and T.
Stanev, Phys. Rev. D {\bf 50}, 5710 (1994).


\bibitem{qgsjet} N. N. Kalmykov, S. S. Ostapchenko, A. I. Pavlov,
Nucl. Phys. B (Proc. Supp.) {\bf B52}, 17 (1997). Details on hadron-nucleus
interactions as described by {\sc qgsjet} are discussed in,
A. B. Kaidalov, K. A. Ter-Martirosyan and Yu. M. Shabel'skii,
Yad. Fiz. {\bf 43}, 1282 (1986) [Sov. J. Nucl. Phys. {\bf 43}, 822 (1986)].

\bibitem{ranft} J. Ranft, [hep-ph/9911232], and [hep-ph/9911213].


\bibitem{prdhi} L. A. Anchordoqui, M. T. Dova, L. N. Epele and
S. J. Sciutto, Phys. Rev. D {\bf 59}, 094003 (1999).



\bibitem{tere} M. T. Dova, D. Heck, J. Knapp and S. J. Sciutto, talk delivered 
at the Auger Collaboration Meeting, Malargue, Argentina, March  2000. 

\bibitem{hillas} A. M. Hillas, in {\it Proc. of the 16$^{\rm th}$
International Cosmic Ray Conference}, Tokyo, Japan, 1979
(University of Tokyo, Tokyo, 1979), Vol.8,p.7.; updated in,
{\it Proc. of the 17$^{\rm th}$ International Cosmic Ray Conference},
Paris, France, 1981 (CEN, Saclay, 1981), Vol.8,p.183.


\bibitem{neu6} G. Cooperman, L. Anchordoqui, V. Grinberg, 
T. McCauley, S. Reucroft and J. Swain, {\it Scalable Parallel 
Implementation of Geant4 Using Commodity Hardware and Task 
Oriented Parallel C}, in Proc. CHEP 2000 (only available in the 
CD version) [hep-ph/0001144].

\bibitem{GEANTREF} http://wwwinfo.cern.ch/asd/geant/

\bibitem{gene} G.~Cooperman, ``{\sc top-c}:  A Task-Oriented Parallel~C
Interface'', {\sl $5^{\hbox{th}}$ International Symposium on High
Performance Distributed Computing} (HPDC-5), IEEE Press,  1996,
pp.~141--150.


\bibitem{atmosphere} {\it U.S. Standard Atmosphere} 1962, updated 1976, 
U.S. Government Printing Office.


\bibitem{auger} http://www.auger.org


\end{thebibliography}
\end{document}